\documentclass[a4paper]{spie}  

\usepackage{amsmath,amsfonts,amssymb}
\usepackage{graphicx}
\usepackage{subcaption}
\usepackage[colorlinks=true, allcolors=blue]{hyperref}
\usepackage{siunitx}

\usepackage{tikz}
\usetikzlibrary{arrows.meta, fit, positioning, backgrounds, calc, shapes.geometric, decorations.pathreplacing}
\usepackage[T1]{fontenc}
\usepackage{listings}
\colorlet{json-string}{black!55}
\colorlet{json-number}{black}
\colorlet{json-bool}{black}

\lstdefinelanguage{json}{
  basicstyle=\ttfamily\footnotesize,
  breaklines=true,
  breakatwhitespace=false,
  columns=fullflexible,
  keepspaces=true,
  frame=single,
  framesep=4pt,
  xleftmargin=4pt,
  xrightmargin=4pt,
  showstringspaces=false,
  stringstyle=\color{json-string},
  string=[b]",
  literate=
    *{0}{{{\color{json-number}0}}}{1}
     {1}{{{\color{json-number}1}}}{1}
     {2}{{{\color{json-number}2}}}{1}
     {3}{{{\color{json-number}3}}}{1}
     {4}{{{\color{json-number}4}}}{1}
     {5}{{{\color{json-number}5}}}{1}
     {6}{{{\color{json-number}6}}}{1}
     {7}{{{\color{json-number}7}}}{1}
     {8}{{{\color{json-number}8}}}{1}
     {9}{{{\color{json-number}9}}}{1}
     {:}{{{\color{black}{:}}}}{1}
     {,}{{{\color{black}{,}}}}{1}
     {true}{{{\color{json-bool}true}}}{4}
     {false}{{{\color{json-bool}false}}}{5}
     {null}{{{\color{json-bool}null}}}{4},
}
\lstdefinestyle{jsonl}{language=json}

\usepackage{xcolor}

\newcommand{\arcsec}{\ensuremath{^{\prime\prime}}}

\title{DUET: Design of a simultaneous InGaAs--CCD split-beam imager for ultra-cool dwarf exoplanet transit survey}

\author[a,b]{Peter P. Pedersen}
\author[a,b]{Clark Baker}
\author[a]{Karolina Dziewiecka}
\author[a]{Mathias Beck}
\author[c]{Michaël Gillon}
\author[d]{Amaury H.M.J. Triaud}
\author[a,b]{Didier Queloz}

\affil[a]{ETH Z\"{u}rich, Department of Physics, Wolfgang-Pauli-Strasse 2, 8093 Zurich, Switzerland}
\affil[b]{Cavendish Laboratory, University of Cambridge, Cambridge, United Kingdom}
\affil[c]{Astrobiology Research Unit, Universit\'{e} de Li\`{e}ge, All\'ee du 6 ao\^ut 19, 4000 Li\`{e}ge, Belgium}
\affil[d]{School of Physics and Astronomy, University of Birmingham, Edgbaston, Birmingham B15 2TT, UK}

\authorinfo{Further author information: (Send correspondence to P.P.P.)\\
P.P.P.: E-mail: ppedersen@phys.ethz.ch}

\begin{document}
\maketitle

\begin{abstract}
We present the design of DUET, a dual-channel imager being built for the SPECULOOS-Southern Observatory (SSO) to detect and characterise transiting terrestrial exoplanets orbiting ultra-cool dwarfs. A dichroic beamsplitter at 955\,nm directs visible light to a deeply-depleted silicon CCD and near-infrared light to a CMOS-based InGaAs detector, enabling fully simultaneous photometry with bandpasses from 0.4 to 1.7\,\textmu m.  The near-infrared filters have been chosen to suppress sensitivity to atmospheric precipitable water vapour (PWV) variability, a dominant source of correlated noise in ground-based infrared photometry of cool stars. We describe the optical and mechanical architecture, the dichroic and filter selection, the choice of detectors, and the control system.
\end{abstract}


\section{INTRODUCTION}
\label{sec:intro}

The Search for habitable Planets EClipsing ULtra-cOOl Stars (SPECULOOS) is a ground-based exoplanet transit survey of the nearest ultra-cool dwarfs (UCDs). 
The SPECULOOS-Southern Observatory (SSO) at ESO Paranal carries out the southern half of this survey.\cite{Gillon2018, delrez2018speculoos, 10.1093/mnras/staa1283, Sebastian20, 10.1117/12.2563563, 10.1117/12.3020550} SSO comprises four robotic 1\,m Ritchey-Chr\'{e}tien telescopes, named Io, Europa, Ganymede, and Callisto. 

The detectability of exoplanets is limited by photometric and systematic noise sources that can mimic or obscure the transit signal. UCDs emit the majority of their flux in the near-infrared, therefore, SPECULOOS is optimised for this regime with deep-depletion CCDs (Andor iKon-L 936 BEX2-DD), and recently with the addition of the SPeculoos InfraRed Imager for Transits (SPIRIT), a CMOS InGaAs-based instrument.\cite{pedersen2024spirit, jano_munoz_spirit} 

\begin{figure}[ht]
  \centering
  \includegraphics[width=\columnwidth]{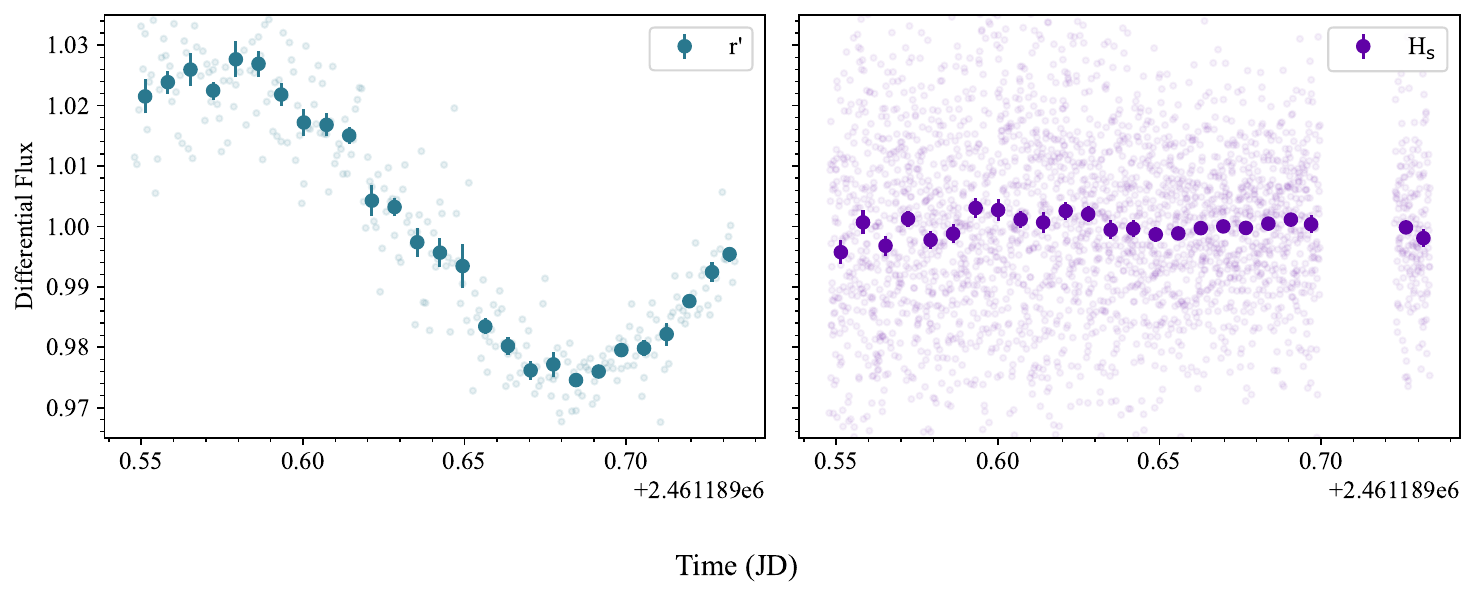}
  \caption{Simultaneous observations of chromatic stellar variability in a mid M-dwarf ($2974\pm100$\,K, M5.5), obtained in the \textit{r'} and $H_s$ bands using SSO Europa and Callisto with exposure times of 48\,s and 5\,s, respectively. The light points show the individual measurements from the raw images, while the larger filled points represent data binned into 10-minute intervals. Error bars indicate the standard deviation within each bin, scaled by the square root of the number of points in that bin. The gap in $H_s$ was due to Callisto's independent weather station triggering a closure alert during robotic operation.}
  \label{fig:variability}
\end{figure}

UCDs are magnetically active; some with detectable rotation periods show peak-to-trough photometric modulations of a few percent at visible wavelengths.\cite{rackham2018tlse} 
It is believed that this variation is primarily driven by cool starspots and hot faculae rotating in and out of view, as exemplified by simultaneous observations in Figure\,\ref{fig:variability} of a mid M-dwarf ($2974\pm100$\,K M5.5) in \textit{r'} and $H_s$, performed by Europa (Andor iKon-L 936) and Callisto (SPIRIT), respectively. 
The variability is strongly chromatic due to the wavelength-dependent contrast of stellar features to the surrounding photosphere.\cite{rackham2018tlse}

The study of chromatic stellar variability has led to the development of DUET, the DUal-band Explorer for Transits. The instrument splits the existing telescope beam at 955\,nm into visible and near-infrared channels, recording photometry simultaneously in two bands.  The near-infrared channel uses custom filters that suppress sensitivity to precipitable water vapour (PWV) variability, a dominant source of correlated noise in ground-based infrared photometry of cool stars.\cite{10.1093/mnras/staa1283, pedersen2023pwv} DUET systems will be installed on two telescopes at SSO. 

The remainder of this paper describes the instrument architecture and optical design of DUET. Section\,\ref{sec:architecture} covers the mechanical layout, detector selection, dichroic and filter optimisation, and optical and photometric performance. Section\,\ref{sec:conclusion} summarises the design and outlines the path to first light.

\section{INSTRUMENT ARCHITECTURE AND OPTICAL DESIGN}
\label{sec:architecture}

\begin{figure}[ht]
  \centering
  \includegraphics[width=\columnwidth]{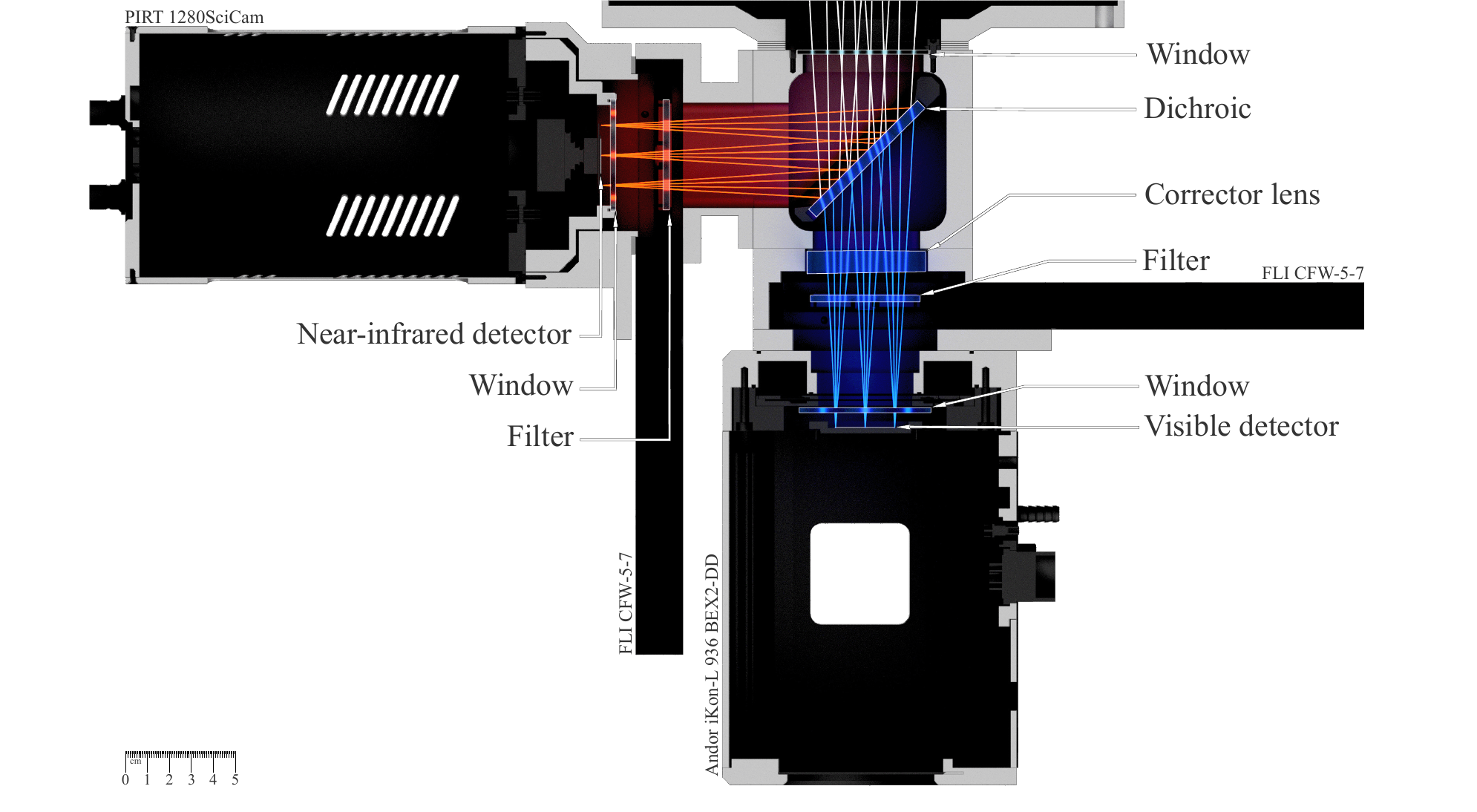}
  \caption{Schematic of the DUET optical path. An F/8 beam (white) from the telescope are split at the dichroic into the near-infrared (red) and visible (blue) arm. Both arms have a 7 position filter wheel (FLI CFW-5-7) holding 50\,mm square filters. The visible arm has a corrector lens before the light reaches the filter and detector.}
  \label{fig:optical_schematic}
\end{figure}

DUET will replace the existing single-camera assembly at two respective telescopes at SSO with a dual-channel layout for each single telescope. Figure\,\ref{fig:optical_schematic} illustrates the optical path. A F/8 beam from the telescope is split into a visible arm and a near-infrared arm, each with its own filter wheel and detector. The visible arm includes a single novel corrector lens to compensate for the aberrations introduced through the plate dichroic, while the infrared arm has no need for correcting optics, as discussed in Section\,\ref{sec:opt_performance}.

The near-infrared arm was chosen to be reflected to allow space for a future expansion of the optical system with a field expansion optic to increase the near-infrared camera's field of view, subsequently increasing the number of comparison stars available for differential photometry, a limiting source of noise in sparse fields. This configuration is led by the lack of space behind the telescope's optical tube assembly (OTA), constrained by the equatorial mount. Reflecting the near-infrared arm also minimises the number of optical elements and simplifies the mechanical design.

For fine alignment of the optical system, the mechanical design aims for a thermally and mechanically stable assembly that uses a combination of stainless steel and plastic shims of various thicknesses, the smallest being 50\,\textmu m thick for the focus matching in both arms. Instead of kinematic mounts, we will use 10\,\textmu m-thick shims that allow adjustment of the fine tip/tilt on the dichroic to account for manufacturing imperfections in matching the image centres between the respective detectors. The instrument envelope, without cabling, covers $335 \times 581 \times 261$\,mm.

Both cameras and filter wheels will be controlled via ASCOM, connected to a dedicated industrial computer (OnLogic Karbon 803) mounted at the back of the OTA, and exposed as ASCOM Alpaca devices for control by SPECULOOS' observatory control software, Astra.\cite{astra_zenodo, pedersen2026astra}

\subsection{Detectors}
\label{sec:detectors}

DUET will use cameras previously utilised at SSO. The visible arm will use the deeply-depleted silicon CCD-based Andor iKon-L 936 BEX2-DD, while the near-infrared arm will use the InP-substrate-removed InGaAs CMOS-based Princeton Infrared Technologies (PIRT) 1280SciCam. The near-infrared camera was characterised and operated as part of the SPIRIT instrument.\cite{pedersen2024spirit}  The visible camera will be air-cooled, whereas the near-infrared camera will be liquid-cooled with a modified Solid State Cooling Systems UC160 TEC Chiller. Table\,\ref{tab:camera-specs} details the specifications for the respective cameras.

\begin{table}[!th]
  \centering
  \caption{Respective camera properties for the visible and near-infrared arm.}
  \begin{tabular}{lrrr}
    \hline
    Parameter & Unit & Visible & Near-infrared\\
    \hline
    Array format & pixel & 2048\,$\times$\,2048 & 1024\,$\times$\,1280 \\
    Pixel pitch & \unit{\um} & 13.5 & 12 \\
    Readout duration & s & 4.5 & 0.1 \\
    Data output & bits & 16 & 14 \\
    Gain & e$^-$/ADU & 1.077 & 5.092 \\
    Well capacity & e$^-$/pixel & 66529 & $\sim$ 56470 \\
    Read noise & e$^-$/pixel & 5.96 & 89.95 \\
    Dark + thermal & e$^-$/s/pixel & 0.2 & $\sim$ 130 \\
    Bad pixels & \% & 0 & 0.325 \\
    Operating temperature & \textdegree C & $-60.0$ & $-50.0$ \\
    Field of view & \unit{'} & 11.9\,$\times$\,11.9 & 5.3\,$\times$\,6.6 \\
    \hline
  \end{tabular} 
  \label{tab:camera-specs}
\end{table}

For dark and bias images, the near-infrared arm's filter wheel will have one 50\,mm square aluminium plate in its filter wheel to block light during calibration acquisitions.  The visible arm will use the camera's integrated shutter.

\subsection{Dichroic and filter selection}
\label{sec:filters}

The dichroic and filter selection were optimised to minimise the effect of PWV variability on the near-infrared arm's photometry, whilst maximising the efficiencies in the respective arms. The respective dichroic and system efficiencies (\textit{g'}, \textit{r'}, \textit{i'}, \textit{z'}, \textit{I+z'} for the visible arm; \textit{Y}, \textit{YJ}, \textit{J}, $H_s$ for the near-infrared arm) are shown in Figure\,\ref{fig:filters}. The observed bandpass cut-ons and cut-offs are listed in Appendix\,{\ref{sec:bandpasses}}.

\begin{figure}[t]
  \centering
  \includegraphics[width=\columnwidth]{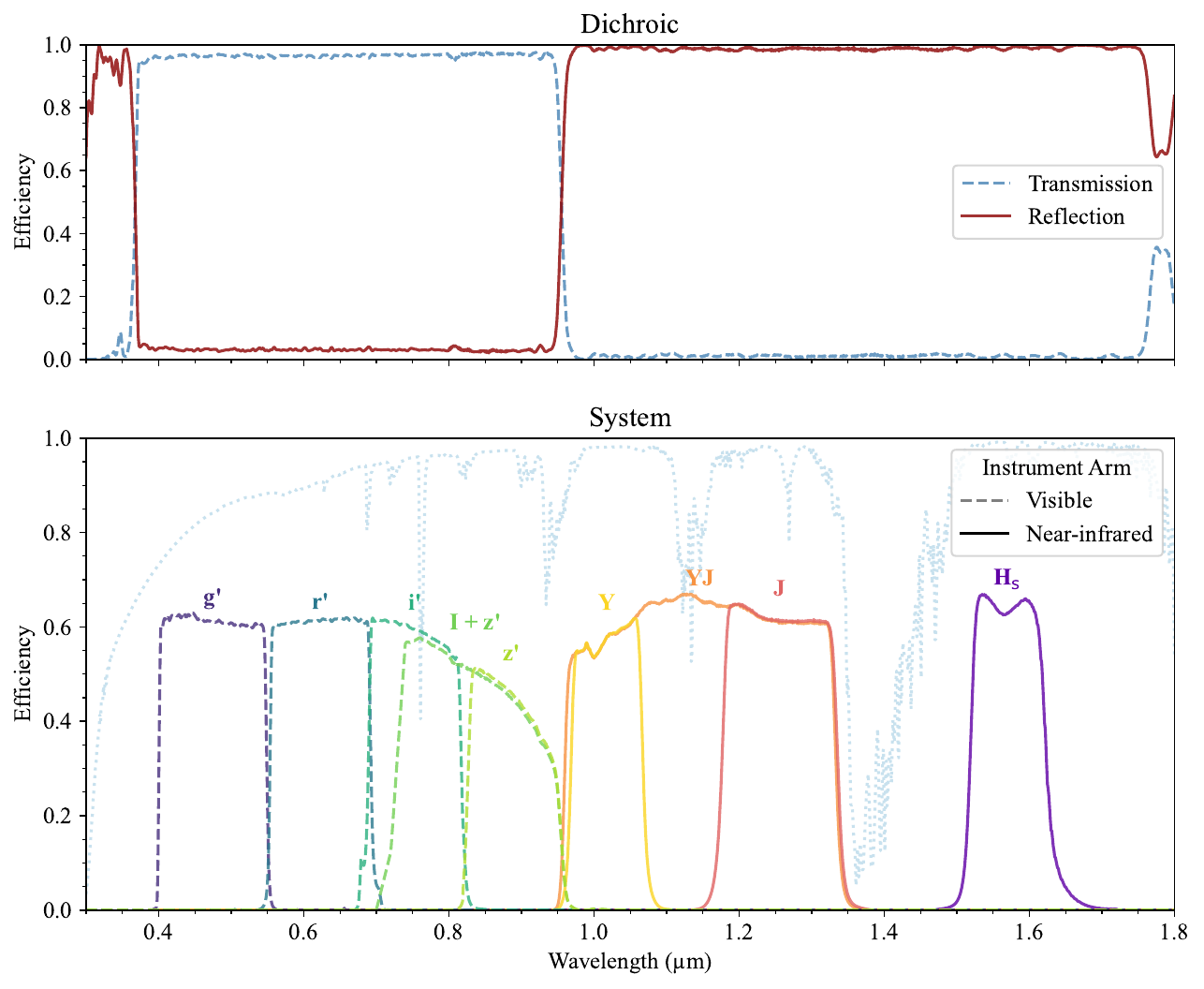}
  \caption{Top: Measured transmission and reflection efficiencies of one of the DUET dichroics.  Bottom: Combined system efficiencies from the telescope, dichroic, filter, and respective detector quantum efficiencies. Superimposed, in dotted light blue, median (airmass of 1, PWV of 2.5\,mm) atmospheric transmission observed in Paranal, Chile.}
  \label{fig:filters}
\end{figure}

The visible filters are standard Sloan filters, previously used by SPECULOOS.\cite{delrez2018speculoos} The near-infrared filters and dichroic are newly acquired from Asahi Spectra. 
Following the work of SPIRIT\cite{pedersen2024spirit}, the wavelength optimisation led to a dichroic transition point of 955\,nm and a custom bandpass of \textit{YJ} and refined \textit{J} and $H_s$. The quantum efficiency cut off of the near-infrared detector of 1.6\,\textmu m dictated the cut off of $H_s$.

\begin{figure}[t]
  \centering
  \includegraphics[width=\columnwidth]{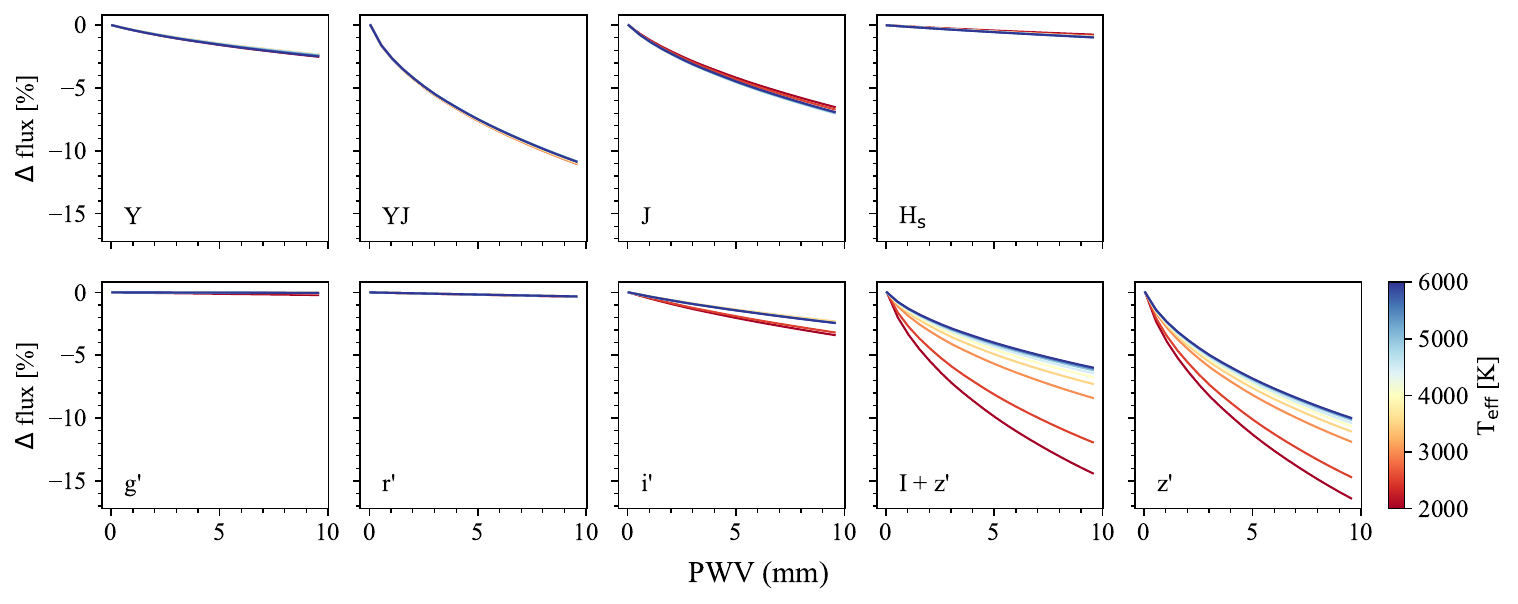}
  \caption{Modelled change in flux as a function of PWV for different-temperature stars (from 6000\,K to 2000\,K in steps of 500\,K) as modelled through the respective bandpasses, with respect to a 0.05\,mm PWV Paranal atmosphere at airmass of 1.}
  \label{fig:flux_vs_pwv}
\end{figure}

The modelled effects of PWV variability on the flux in each bandpass are shown in Figure\,\ref{fig:flux_vs_pwv}. The modelling demonstrates the reduced second order variability of the change of flux as a function of PWV and effective stellar temperature in the near-infrared bandpasses, as indicated by the minimised spread of the change in flux for different stellar temperatures for increasing PWV. The visible arm's filters were not optimised as part of this work.

\subsection{Optical performance}
\label{sec:opt_performance}

The SPECULOOS telescopes are 1\,m aperture Ritchey-Chrétien telescopes that image at F/8 to an approximately $12' \times 12'$ square field of view. Optically, they are designed with the capacity to achieve diffraction-limited performance. However, being a ground-based observatory, they operate in a seeing-limited capacity.

A concept study was undertaken for the optical design of DUET, which found that the two aberration correcting lenses that the telescopes have close to their focal plane were unnecessary given the seeing conditions of the site (approximately $0.8\unit{\arcsec} \rightarrow 1.3\unit{\arcsec}$, equating to $31 \rightarrow 50$\,\unit{\um} on the detector plane) and the field of view used by the SPECULOOS visible and near-infrared cameras (see Table\,\ref{tab:camera-specs}). Without these corrector lenses, the optical aberration performance of the telescope is seen to generate spots with 80\% of the enclosed energy (EE80) within a diameter of 30\,--\,44\,\unit{\um} across the majority of the field. This greatly benefits the optical design of DUET, opening additional mechanical envelope for the dichroic system from the telescope's interface plate to its ground clearance. 

Initially, a \textit{cube-type} dichroic beam-splitter was pursued to separate the two arms of this system. This was desired as the flat, and perpendicular to the optical axis, faces of the cube would induce minimal and similar aberrations for both the visible and infrared arms of the DUET system. However, due to the sharp cut-on/off wavelength transition and high efficiency requirements of the instrument, it was found that the number of dielectric coating layers (and thickness of the resultant layers) required to produce the necessary system performance was not compatible with a \textit{cube-type} beam-splitter. Hence, a \textit{plate} dichroic was investigated instead.

Once the \textit{cube-type} beam-splitter was replaced with a \textit{plate} dichroic at $45^{\circ}$ to the optical axis, while there was a negligible change in optical performance for the reflected infrared arm, there was an expected and substantial degradation in imaging performance of the transmitted visible arm. A combination of introduced field-curvature, astigmatism, and spherical aberration brought EE80 performance to $\approx160$\,\unit{\um}; far beyond the performance required by the system ($<45$\,\unit{\um}).

\begin{figure}
    \centering
    \includegraphics[width=0.5\linewidth]{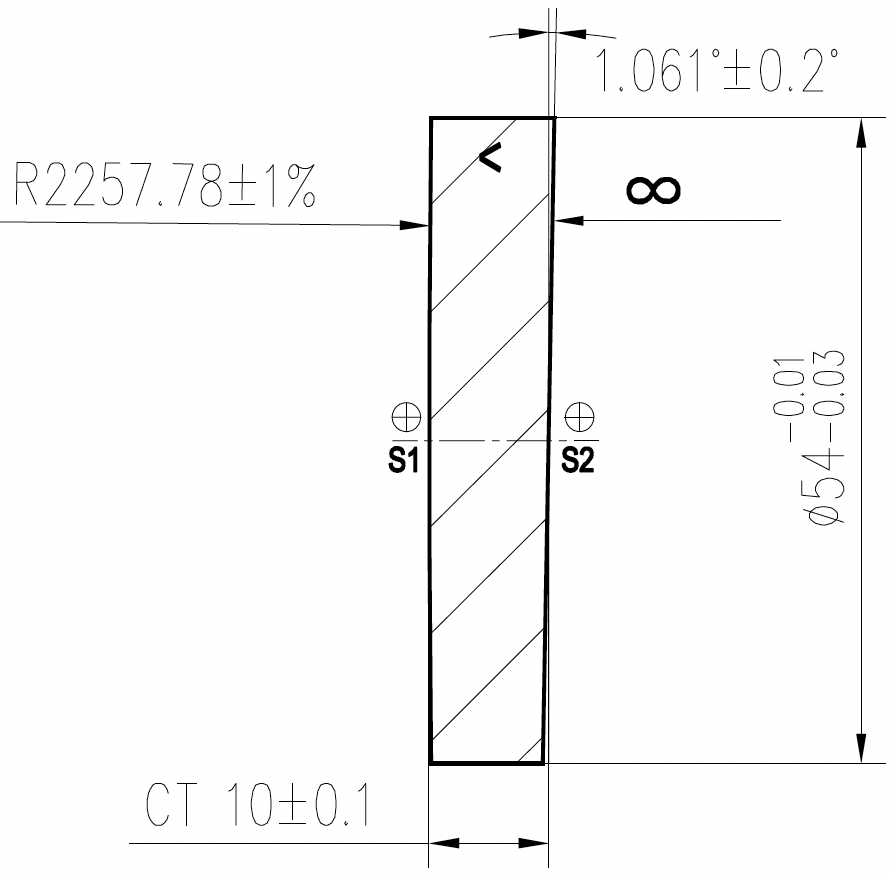}
    \caption{Schematic drawing of the cylindrical wedged field lens. `R' notes the radius and tolerance on the cylinder of surface 1 (S1), `CT' notes the central thickness of the lens and, labelled on surface 2 (S2), is the wedge angle and lens diameter. All length measurements are in mm.}
    \label{fig:lens}
\end{figure}

To combat these aberrations, a novel correction solution was devised. Instead of using a standard multi-element approach, a single cylindrical wedged field lens was designed (shown in Figure\,\ref{fig:lens}). This enabled the required correction within the space constraints of the optical path (12\,mm) and simultaneously maximised throughput. Surface 1 is a large radius cylinder, which primarily corrects the plate-induced astigmatism and spherical aberration, and surface 2 is a shallow wedge about the cylindrical axis, which allows the correction of field curvature and astigmatism. Beyond this, the lens was further optimised to correct for the majority of aberrations that were previously corrected by the two original corrector lenses. The lens is composed of an N-BK7 equivalent glass (H-K9L) and has tuned dielectric coatings on both surfaces for our visible waveband of 400\,--\,955\,nm. The prescription of this lens is noted in Figure\,\ref{fig:lens}. Owing to the design approach taken, which minimised break-over angles of the optical path at the two surfaces, and through being located close to the focal plane, where the marginal rays are close to the principle rays, this lens is extremely insensitive to position; with close to nominal performance achieved with extremely loose positional tolerances of a few millimetres and tip/tilt tolerance on the order of several degrees.

The optical windows and filters are manufactured from fused silica and are wedged to reduce spectral fringing. The dichroic substrate is similarly constructed of SK-1300 fused silica with a 45" wedge.

\begin{figure}[htbp]
    \centering
    \begin{subfigure}[b]{0.8\textwidth}
        \centering
        \includegraphics[width=\linewidth]{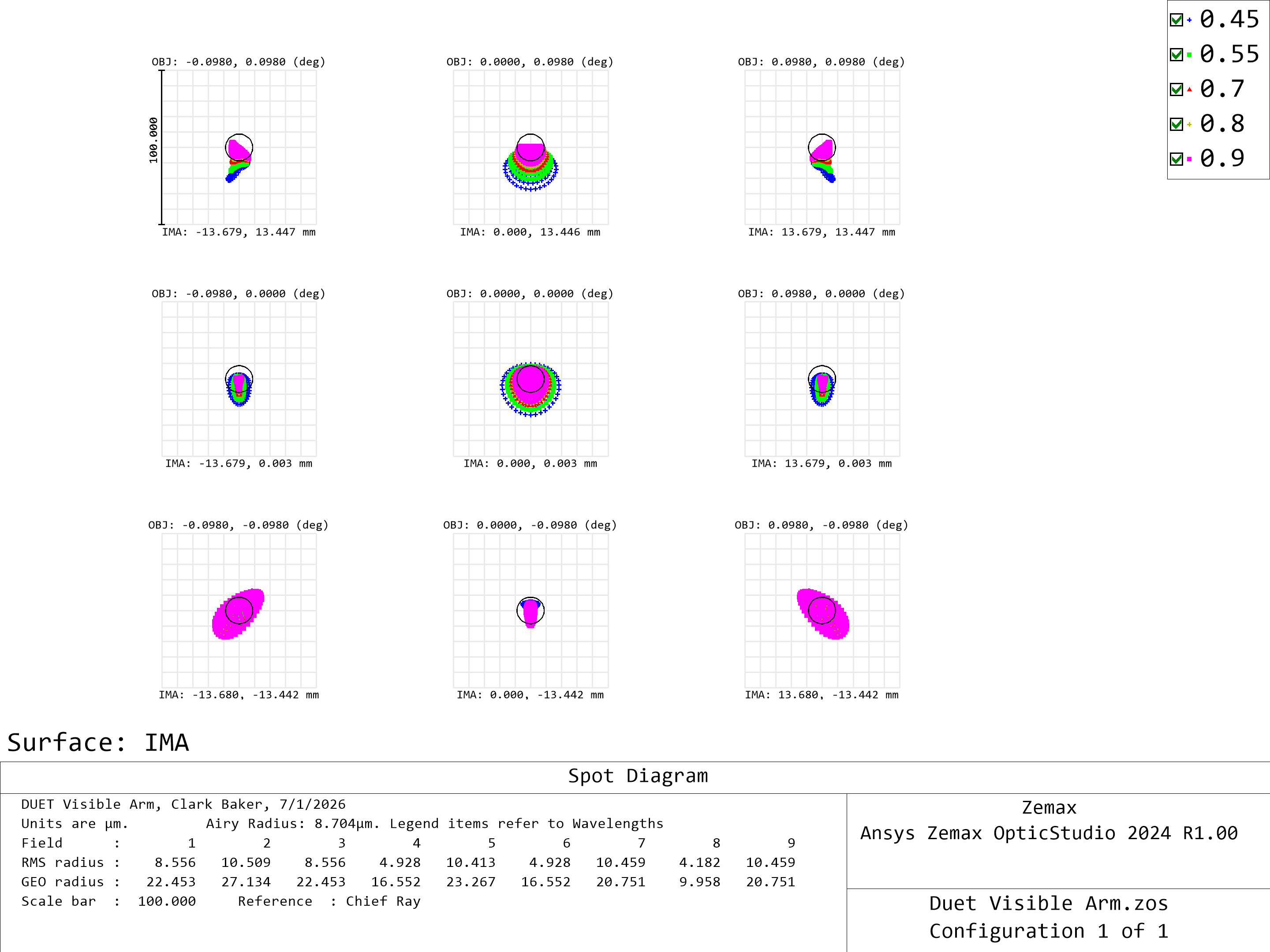}
        \caption{}
        \label{fig:first}
    \end{subfigure}
    
    \vspace{0.5em} 
    
    \begin{subfigure}[b]{0.8\textwidth}
        \centering
        \includegraphics[width=\linewidth]{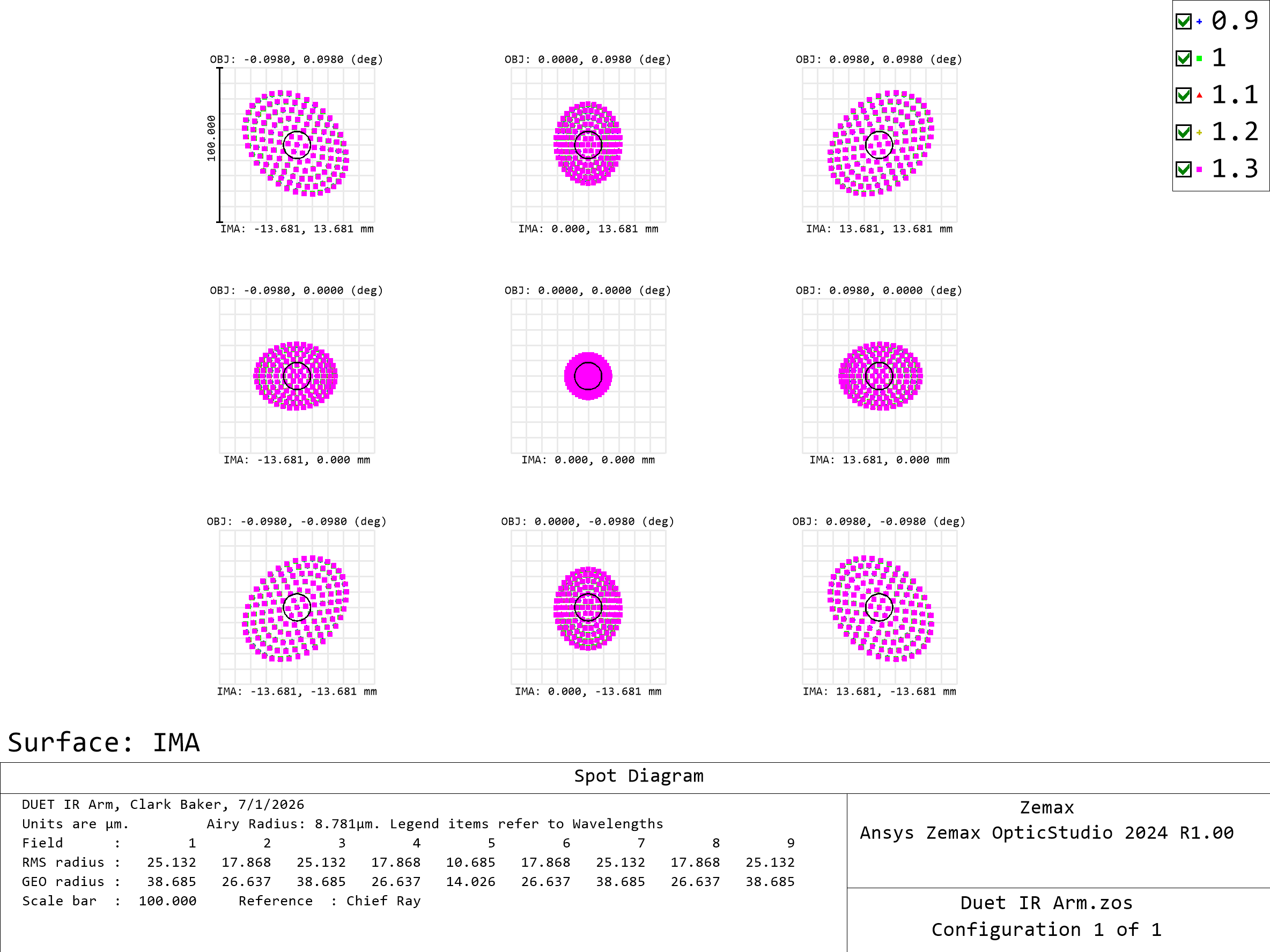}
        \caption{}
        \label{fig:second}
    \end{subfigure}
    
  \caption{Zemax spot diagrams for the DUET visible arm (a, 0.45--0.9\,\textmu m) and near-infrared arm (b, 0.9--1.3\,\textmu m), sampled at the centre, edges, and corners of the field ($\pm0.098^\circ\approx6'$). Black circles denote the respective theoretical Airy disk radii (8.70\,\textmu m and 10.73\,\textmu m). The corrected visible arm keeps aberrations well-controlled with a maximum RMS spot radius of 10.51\,\textmu m. The uncorrected infrared arm is diffraction-limited on-axis (10.68\,\textmu m RMS) but degrades to 25.13\,\textmu m at the extreme corners. Crucially, the majority of fields achieve EE80 performance better than 44\,\unit{\um}, which equates to approximately 1.1" on sky; near median seeing of the site.}
  \label{fig:spots}
\end{figure}

\begin{figure}[htbp]
    \centering
    \begin{subfigure}[b]{0.8\textwidth}
        \centering
        \includegraphics[width=\linewidth]{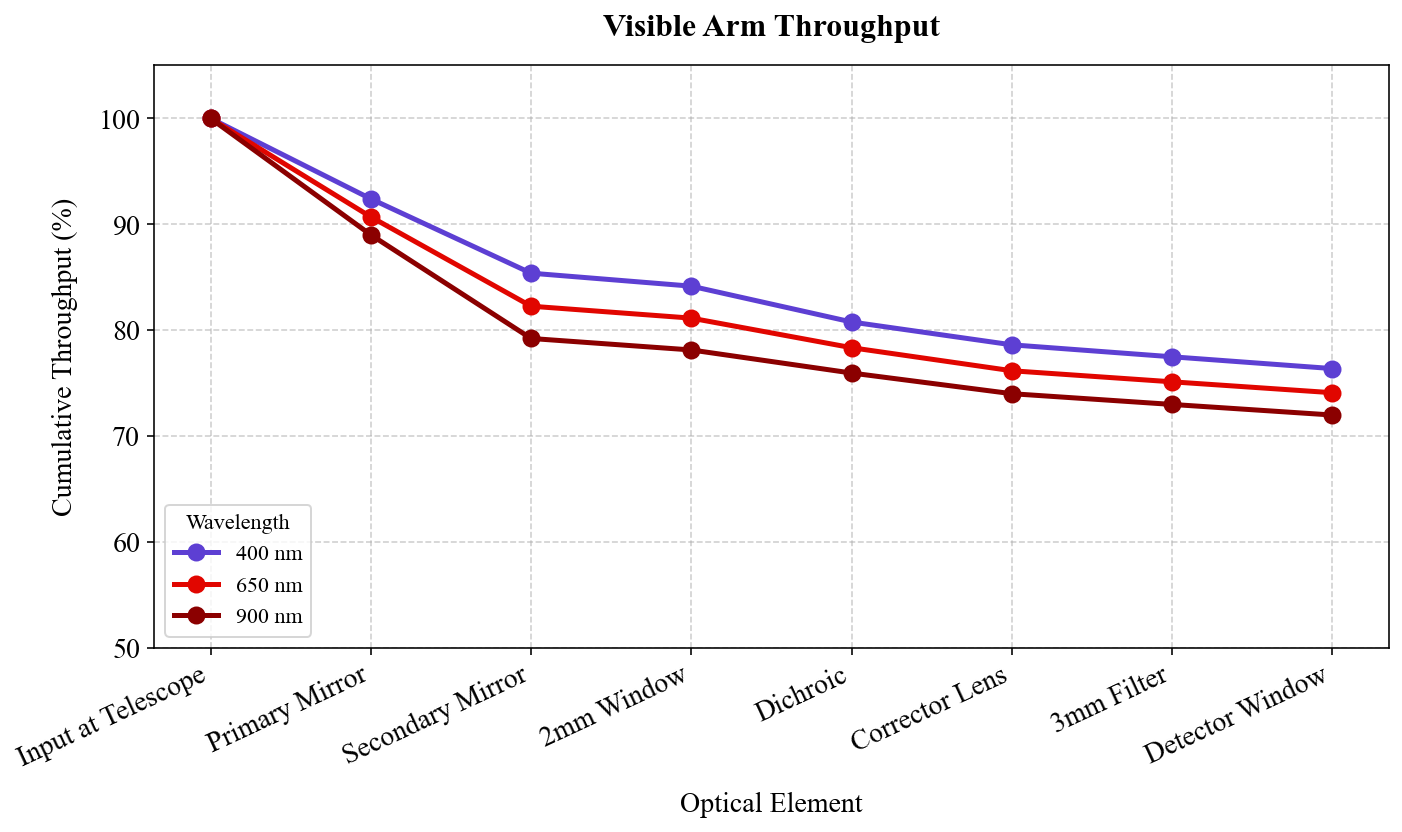}
        \caption{Cumulative optical efficiency plot of the optical components in the DUET visible arm, from the telescope to the detector surface.}
        \label{fig:first}
    \end{subfigure}
    
    \vspace{0.5em} 
    
    \begin{subfigure}[b]{0.8\textwidth}
        \centering
        \includegraphics[width=\linewidth]{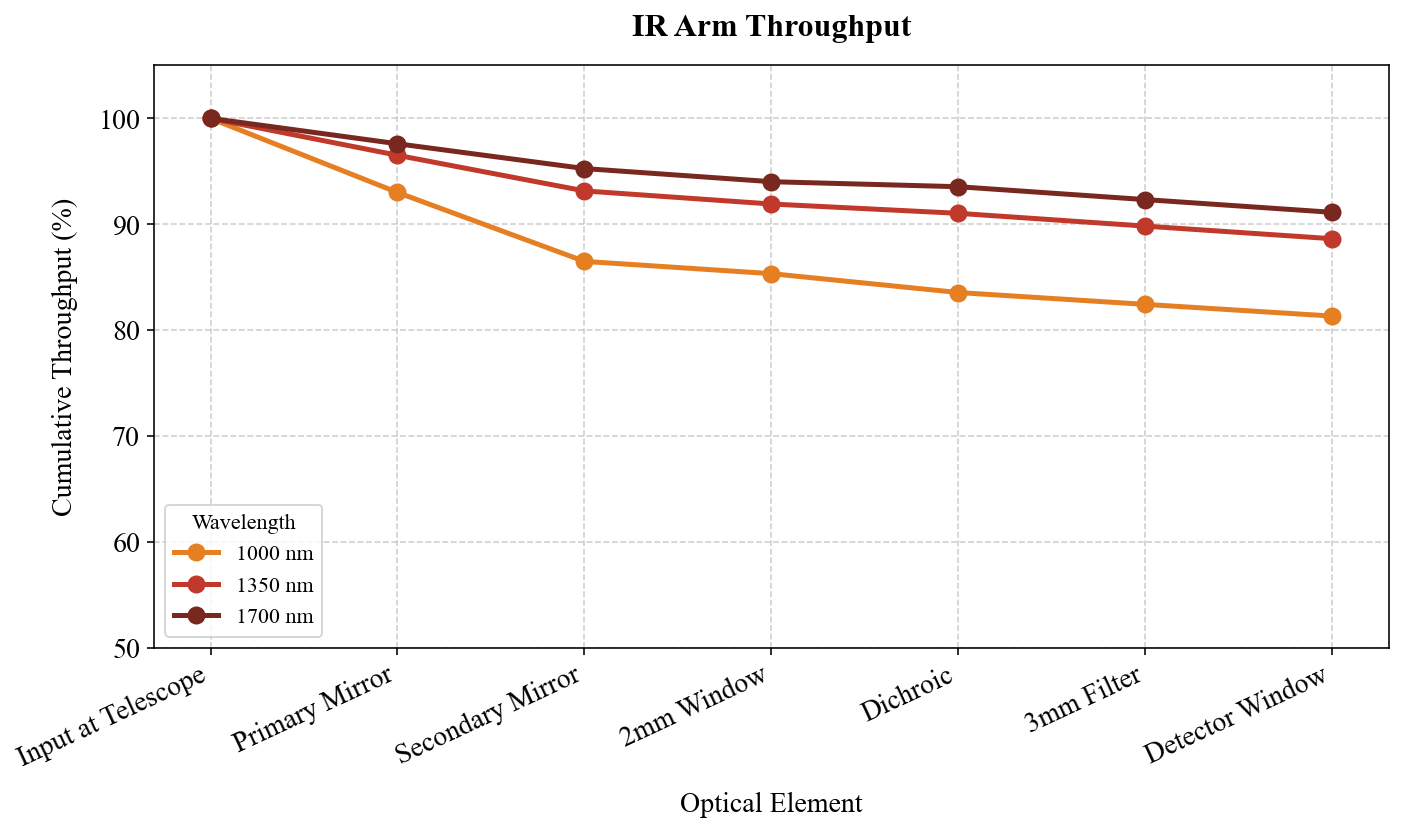}
        \caption{Cumulative optical efficiency plot of the optical components in the DUET infrared arm, from the telescope to the detector surface.}
        \label{fig:second}
    \end{subfigure}
    
    \caption{Waterfall throughput degradation plots for (a) the visible arm and (b) the infrared arm of DUET. The coatings of the bandpass filters are not included here.}
    \label{fig:eff}
\end{figure}

The resultant optical imaging performance of the visible and infrared arms of DUET are detailed in Figure \ref{fig:spots}, with transmission of the SPECULOOS/DUET system demonstrated in Figure \ref{fig:eff}. The visible arm, with its new corrector lens, achieves close to diffraction-limited performance across the entire field of view, while the uncorrected infrared arm achieves sufficient imaging performance to meet the requirements of the photometric science case.

\subsection{Photometric performance}
\label{sec:photometry}

\begin{figure}[ht]
  \centering
  \includegraphics[width=\columnwidth]{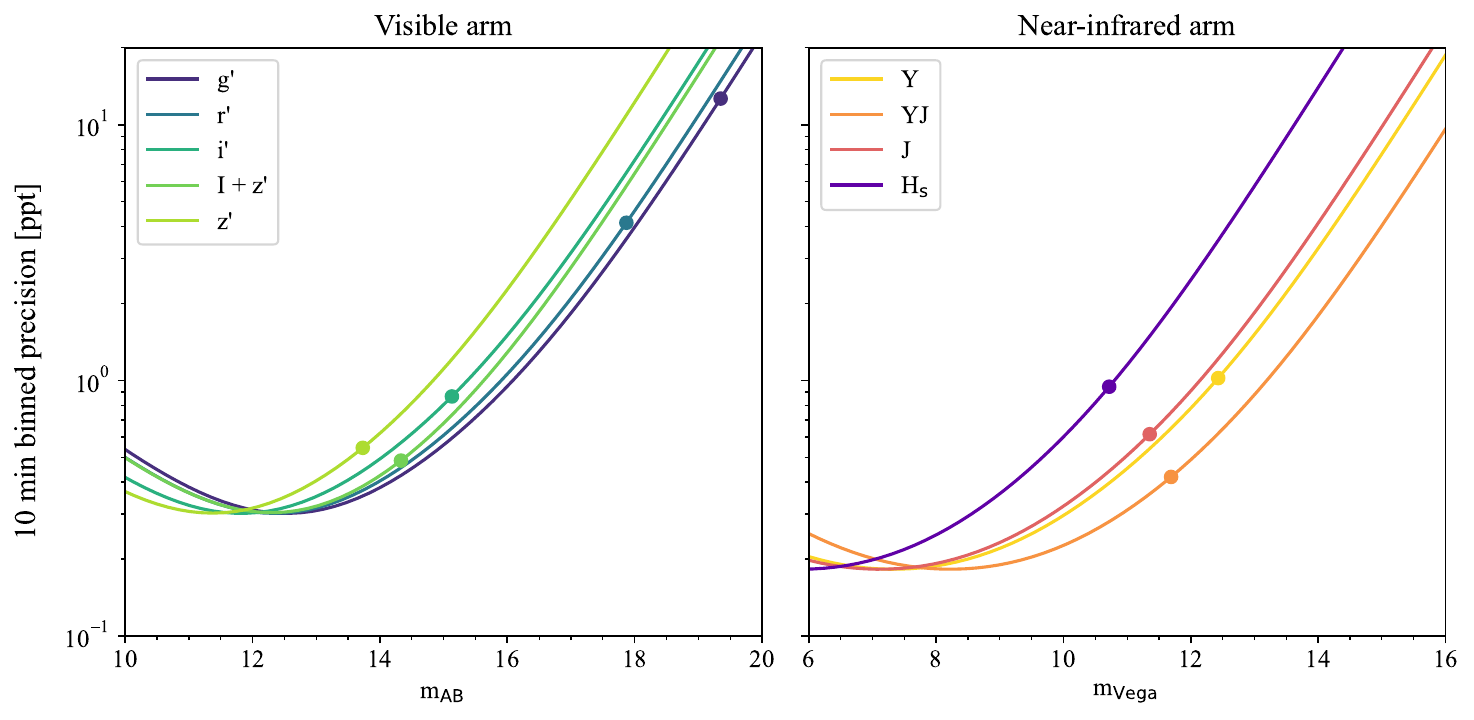}
  \caption{Modelled photometric 10-minute binned precision limiting magnitudes for the respective bandpasses and arms for a TRAPPIST-1-like star under median atmospheric conditions at SSO (airmass=1, PWV=2.5\,mm, seeing=1.35", photometry aperture radius of $3\times$ the seeing). The solid dots are measured magnitudes for the TRAPPIST-1 system from the Pan-STARRS1 (AB) and 2MASS (Vega) survey.\cite{chambers2019panstarrs1surveys, Skrutskie_2006} Y magnitude from Pan-STARRS1 was converted into Vega magnitudes to match 2MASS' native Vega magnitudes.}
  \label{fig:precision}
\end{figure}

End-to-end precision modelling, performed with the \texttt{mphot}\footnote{\url{https://github.com/ppp-one/mphot}} package developed for SPIRIT,\cite{pedersen2024spirit} was used to model the photometric performance of each arm as a function of magnitude. 
Figure\,\ref{fig:precision} shows the 10-minute binned precision-limiting magnitudes for the respective filters, as modelled for a TRAPPIST-1-like star (2556\,K, M8V)\cite{Agol2021}, with the precision-magnitude relationship shifting depending on stellar type.

As Figure\,\ref{fig:precision} shows, the observed photometric precisions will naturally depend on the filters selected. Thus, observing strategy of DUET's targets will be optimised for stellar type, activity scale, and magnitude, to maximise our understanding of stellar properties and the likelihood of disentangling transit signals from activity.
Additionally, any false activity from atmospheric PWV variability, primarily in \textit{I+z'} and \textit{z'}, will need to be mitigated post observations using PWV radiometer data from the nearby LHATPRO.\cite{pedersen2023pwv}

\section{Conclusion}
\label{sec:conclusion}

We have presented DUET, a dual-channel imager designed for the SPECULOOS-Southern Observatory. A dichroic beamsplitter at 955\,nm directs visible light to a deeply-depleted CCD and near-infrared light to an InGaAs CMOS detector, enabling simultaneous photometry from 0.4--1.7\,\textmu m. The instrument is motivated by the chromatic nature of UCD stellar variability and the need to disentangle stellar activity from planetary transit signals, as demonstrated by the simultaneous multi-band observations in Figure\,\ref{fig:variability}.

The optical design keeps both arms within the seeing envelope at Paranal. A single corrector lens in the visible arm compensates for dichroic-induced aberrations, while the uncorrected near-infrared arm remains diffraction-limited on-axis. Near-infrared filters were designed to suppress sensitivity to PWV variability, building on the methodology developed for SPIRIT. Modelled photometric performance confirms that filter selection will require optimisation per target as a function of stellar type, activity level, and magnitude.

Two DUET systems will be installed on two telescopes at SSO, extending the observatory's simultaneous multi-band photometric capability and providing the chromatic baseline needed to characterise stellar activity and recover transit signals.

\newpage
\appendix
\section{Bandpasses}
\label{sec:bandpasses}

\begin{table}[h]
  \caption{50\% cut-on and off points of the respective bandpasses as shown in Figure\,\ref{fig:filters}, measured from the half-maximum.} 
  \label{tab:filter-cuts}
  \begin{center}       
    \begin{tabular}{lrr}
      \hline
      Name & Cut-on [\unit{\um}] & Cut-off [\unit{\um}] \\ 
      \hline
      \textit{g'} & 0.400 & 0.549 \\
      \textit{r'} & 0.554 & 0.693 \\
      \textit{i'} & 0.689 & 0.817 \\
      \textit{I+z'} & 0.727 & 0.947 \\
      \textit{z'} & 0.826 & 0.950 \\
      \hline
      \textit{Y} & 0.968 & 1.068 \\
      \textit{YJ} & 0.961 & 1.333 \\
      \textit{J} & 1.177 & 1.336 \\
      $H_s$ & 1.519 & 1.623 \\
      \hline
      \end{tabular}
  \end{center}
\end{table}

\acknowledgments
The authors are very grateful to Catriona A. Murray, Clàudia Janó-Muñoz, and Sebasti\'{a}n Z\'{u}\~{n}iga-Fern\'{a}ndez for reviewing the manuscript and helping to improve its clarity.
The ULiege's contribution to SPECULOOS has received funding from the European Research Council under the European Union's Seventh Framework Programme (FP/2007-2013) (grant Agreement n$^\circ$ 336480/SPECULOOS), from the Balzan Prize and Francqui Foundations, from the Belgian Scientific Research Foundation (F.R.S.-FNRS; grant n$^\circ$ T.0109.20), from the University of Liege, and from the ARC grant for Concerted Research Actions financed by the Wallonia-Brussels Federation. MG is F.R.S-FNRS Research Director. 
This research is supported by the Science and Technology Facilities Council (STFC; grant n$^\circ$ ST/S00193X/1, ST/W002582/1, and ST/Y001710/1).

\bibliography{report} 
\bibliographystyle{spiebib} 

\end{document}